# Ordering of hidden multipoles in spin-orbital entangled $5d^1$ Ta chlorides

H. Ishikawa,[1,2] T. Takayama,[1,2] R. K. Kremer,[2] J. Nuss,[2] R. Dinnebier,[2] K. Kitagawa,[3] K. Ishii,[4] and H. Takagi[1,2,3]

[1] *Institute for Functional Matter and Quantum Technologies, University of Stuttgart, Stuttgart 70569, Germany*

[2] *Max-Planck-Institute for Solid State Research, Heisenbergstraße 1, Stuttgart 70569, Germany*

[3] *Department of Physics, The University of Tokyo, 7-3-1 Hongo, Bunkyo-ku, Tokyo 113-0033, Japan*

[4] *Synchrotron Radiation Research Center, National Institutes for Quantum and Radiological Science and Technology, Sayo, Hyogo 679-5148, Japan*

Spin-orbit coupling of as large as a half eV for electrons in $5d$ orbitals often gives rise to the formation of spin-orbital entangled objects, with the effective total angular momentum $J_{\text{eff}}$. Of particular interest are the $J_{\text{eff}} = 3/2$ states realized in $5d^1$ transition metal ions surrounded by an anion octahedron. The pure $J_{\text{eff}} = 3/2$ quartet does *not* have any magnetic dipolar moment ($<M> = 0$) but hosts hidden pseudo-dipolar moments accompanied by charge quadrupoles and magnetic octupoles. $Cs_2TaCl_6$ and $Rb_2TaCl_6$ are correlated insulators with $5d^1$ $Ta^{4+}$ ions in a regular Cl octahedron. Here we demonstrate that these Ta chlorides indeed have a substantially suppressed effective magnetic dipolar moment of ~ 0.2 $\mu_B$. Two phase transitions are observed at low temperatures that are not pronounced in the magnetization but accompanied with large electronic entropy of ~ $R\ln 4$. We ascribe the two transitions to the ordering of hidden multipoles.

## I. INTRODUCTION

The spin-orbit coupling for $5d$-electrons in heavy transition metal oxides, up to ~ 0.5 eV, is comparable to other relevant energy scales, such as the crystal-field splitting. This gives rise to a strong entanglement of spin and orbital moments and the formation of spin-orbital-entangled objects, often described by an effective total angular moment $J_{\text{eff}}$. The exploration of novel phases formed by such spin-orbital entangled objects through exchange and other interactions is attracting considerable interest, and $5d$ complex oxides with only $t_{2g}$ electrons are particularly promising. The expected phases include a topological quantum liquid [1], a two-dimensional Heisenberg magnet [2] and an excitonic magnet [3]. A particularly well-studied class of such materials is the $Ir^{4+}$ perovskite-related oxides with five $t_{2g}$ electrons, in which the spin-orbital entangled $J_{\text{eff}} = 1/2$ doublet behaves like an $S = 1/2$ moment.

However, the underlying orbital degrees of freedom often give rise to non-trivial exchange couplings between $J_{eff} = 1/2$ moments, resulting in unusual magnetic states.

Another interesting but not yet fully explored set of materials are the $5d^1$ electron systems, where the $t_{2g}$ electron carries the spin angular momentum $S = 1/2$ and *effective* orbital angular momentum $L_{eff} = 1$. The strong spin-orbit coupling stabilizes the $J_{eff} = 3/2$ quartet against the $J_{eff} = 1/2$ doublet [4] and the one electron occupies the degenerate $J_{eff} = 3/2$ quartet. The $J_{eff} = 3/2$ quartet has an interesting property as $J_{eff}$ and $L_{eff}$ are not a real but an effective momentum; the expectation value of the magnetization $<M> = <2S - L_{eff}>$ for the ideal $J_{eff} = 3/2$ state is *zero* as a result of the cancellation of $L_{eff} = 1$ with $S = 1/2$, resulting in Landé g-factor $g = 0$ [4]. The degeneracy of the $J_{eff} = 3/2$ quartet without magnetic dipolar moments, however, is still characterized by degenerate pseudo-dipolar moments $J_{eff}^z = \pm 3/2$ and $\pm 1/2$, which are accompanied by charge quadrupolar and magnetic octupolar moments [5]. The $J_{eff}^z = \pm 3/2$ state consists of $d_{yz}$ and $d_{zx}$ orbitals, whereas the $d_{xy}$ orbital is dominant in the $J_{eff}^z = \pm 1/2$ state [6]. With strong Coulomb correlation, a magnetically silent $J_{eff} = 3/2$ Mott insulator, with significantly suppressed magnetic dipole moment, can be realized. In contrast to the conventional Mott insulators with large magnetic dipole moments, interactions among higher-order multipolar moments may dominate the physics of $J_{eff} = 3/2$ Mott insulator, which can lead to multipolar ordering [5,6], or, if sufficiently symmetric, to a spin-orbital liquid state [7,8].

The complete degeneracy of the $J_{eff} = 3/2$ quartet is realized when $5d^1$ ions such as $Os^{7+}$, $Re^{6+}$, $W^{5+}$ and $Ta^{4+}$ are placed in highly symmetric and regular oxygen octahedra, as in the case for the cubic double perovskite oxides $A_2BB'O_6$. In cubic $A_2BB'O_6$, the transition metal $B$ ions are placed at the centers of regular oxygen octahedra and form a face-centered-cubic (FCC) sub-lattice. The magnetic properties of $5d^1$ double perovskite oxides such as $Ba_2NaOsO_6$ [9,10] and $Ba_2MgReO_6$ [11,12] have been investigated experimentally. A Curie-Weiss effective magnetic moment ($p_{eff}$) of 0.7-1.7 $\mu_B$ are observed in the magnetic susceptibility, $\chi(T)$, which is not very small, but less than the 1.7 $\mu_B$ expected for a pure $S = 1/2$ moment. The reduced size of the magnetic moment is consistent with the partial cancellation of the spin moment by the orbital moment. The incomplete cancellation can be ascribed to a strong hybridization of $5d$ states with O-$2p$ states [5]. A spinel chalcogenide $GaTa_4Se_8$ may host a molecular $J_{eff} = 3/2$ state on a $Ta_4$ cluster [13, 14]. $GaTa_4Se_8$ has a $p_{eff}$ of 0.86 $\mu_B$ per $Ta_4$ cluster [15], which is comparable to those of $5d^1$ double perovskite oxides and indicative of the strong hybridization between Ta-$5d$ and Se-$4p$ states.

The combination of orbital-dependent exchange interactions with a geometrically frustrated FCC

lattice is believed to give rise to a rich variety of ground states with ordering of pseudo-dipolar and hence multipolar moments [5,6]. The large and uncompensated magnetic dipolar moments in $5d^1$ double perovskite oxides [9-12] and GaTa$_4$Se$_8$ [15] show an ordering or a spin-glass freezing at a low temperature. For example, Ba$_2$NaOsO$_6$ exhibits a weak ferromagnetic ordering with a saturation moment of 0.2 $\mu_B$/Os and (110) easy-axis [9]. A recent NMR study indicates that the in-plane moments are canted around the (110) axis in the ferromagnetic phase, with an angle close to that expected for the (110) ferromagnetic phase proposed theoretically as a ground state of interacting $J_{\text{eff}} = 3/2$ pseudo-dipolar moments [5]. An electric-quadrupolar ordering is predicted to occur above the ferromagnetic ordering in the presence of a sizable quadrupolar interaction [5]. In the systems investigated so far, the signature of such quadrupolar ordering is very weak and its presence is still a subject of debate. The change of the NMR spectra in Ba$_2$NaOsO$_6$ [10] and the presence of specific heat anomaly in Ba$_2$MgReO$_6$ [12] above the magnetic ordering temperatures were suggested to represent the quadrupolar ordering. Any structural phase transition induced by the quadrupolar ordering, however, has not been identified in the two systems and in the other candidates of $J_{\text{eff}} = 3/2$ systems.

We have been exploring new $5d^1$ compounds that clearly show the effect of strong spin-orbital entanglement, for example by an almost complete suppression of the magnetic dipolar moment and the existence of multipolar ordering. Our strategy was to reduce the $d$-$p$ hybridization, which drew our attention to the chloride family $A_2$TaCl$_6$ ($A$ = Cs, Rb, K) [16-18]. These materials have a cubic crystal structure closely related to that of double perovskite oxides $A_2BB'$O$_6$ at room temperature. By writing $A_2$TaCl$_6$ as $A_2$Ta□Cl$_6$, it is clear that $B'$ in $A_2BB'$O$_6$ is a cation vacancy □ and that O is replaced with Cl. $5d^1$ Ta$^{4+}$ ions are at the center of regular and highly symmetric Cl$_6$-octahedra and form an FCC sub-lattice as in double perovskite oxides (Fig. 1(a)).

In this paper, we show that the magnetic dipolar moment of $5d^1$ electrons in $A_2$TaCl$_6$ ($A$ = Cs, Rb, K) is almost completely suppressed, as expected for the pure $J_{\text{eff}} = 3/2$ state. In Cs$_2$TaCl$_6$ and Rb$_2$TaCl$_6$, two consecutive transitions are observed with only a very weak magnetic signature but with a large total entropy change close to $R$ln4. The high-temperature transition is accompanied by a tetragonal distortion of the TaCl$_6$ octahedra. We ascribe the high-temperature transition to a hidden ordering of charge quadrupoles and the low-temperature transition to an ordering of pseudo-dipolar moments with magnetic octupoles with substantially suppressed magnetic dipoles.

## II. EXPERIMENAL METHODS

## A. Sample preparation

Powder samples of $A_2$TaCl$_6$ ($A$ = K, Rb, Cs), Rb$_2$NbCl$_6$ and Cs$_2$HfCl$_6$ were synthesized by a solid-state reaction. The stoichiometric mixture of $A$Cl ($A$ = Cs, Rb and K), Ta, TaCl$_5$, Nb, NbCl$_5$, HfCl$_4$ powders in an evacuated quartz tube was heated at 823 K for 96 h. The starting chemicals and products are air sensitive. All the synthetic processes were performed in an Ar-atmosphere. Single crystals of K$_2$TaCl$_6$ were obtained from 2 : 1 mixture of KCl and TaCl$_5$ powders. The mixture was placed in an evacuated quartz tube with the tantalum metal wire. The tube was first heated to 873 K and then slowly cooled down to 773 K in 100 h.

## B. Powder x-ray diffraction measurements and structural analysis

Powder X-ray diffraction (XRD) measurements of $A_2$TaCl$_6$ ($A$ = Cs, Rb, K) were conducted at various temperatures using two laboratory diffractometers. For Rb$_2$TaCl$_6$, Bruker D-8 Discover system was used, with Cu-$K\alpha_1$ radiation from primary Ge(111) Johannson monochromator and VANTEC position sensitive detector in Bragg-Brentano mode. A closed cycle helium cryostat (Phenix, Oxford Cryo-systems) was used to change temperature of the sample. Powder diffraction data were taken from $2\theta$ = 13.0° to 140.0° with steps of 0.016°. For Cs$_2$TaCl$_6$ and K$_2$TaCl$_6$, a Bruker D-8 Advance system was used, with Mo-$K\alpha_1$ radiation from a primary Ge(111) Johannson monochromator and LYNXEYE position sensitive detector in Debye-Scherrer mode. A home-built low-temperature cryostat allows the samples to cool down to 3.8 K [19]. Powder diffraction data were taken from $2\theta$ = 5.0° to 95.0° with steps of 0.004°. For the analysis of powder diffraction patterns, the program TOPAS (Version 5, Bruker AXS) or FullProf [20] was used.

## C. Single crystal x-ray diffraction measurements and structure determination

Crystals of K$_2$TaCl$_6$ suitable for single crystal X-ray diffraction were mounted with some grease on a loop made of Kapton foil (Micromounts, MiTeGen). Diffraction data were collected at 298 and 100 K with a SMART APEX-I CCD X-ray diffractometer (Bruker AXS) equipped with a Cryostream 700 Plus cooling device (Oxford Cryo-systems). The collection and reduction of data were carried out with the Bruker Suite software package [21]. The intensities were corrected for absorption effects applying a multi-scan method with Sadabs [22] or Twinabs [23]. The structure was solved by Direct methods and refined by full matrix least-squares fitting with the Shelxtl software package [24,25]. Experimental details of data collection and crystallographic data are given in the Supplementary Material. Further details can be obtained from Fachinformationszentrum Karlsruhe, 76344 Eggenstein-Leopoldshafen, Germany, on quoting the CSD numbers: CSD-434040 (100 K), and CSD-434041 (298 K).

### D. Resonant inelastic x-ray scattering

The resonant inelastic x-ray scattering (RIXS) measurement at Ta $L_3$-edge was performed on $A_2$TaCl$_6$ ($A$ = Rb and K) powder samples at BL11XU SPring-8. The incident x-ray was monochromatized by a double-crystal of Si(111) monochromator and by a secondary 4-bounce Si(333) asymmetric monochromator. The scattering x-ray was analyzed by Ge(840) diced and spherically-bent analyzer and collected by Mythen microstrip x-ray detector (Dectris). The total energy resolution was ~ 90 meV. Since the $A_2$TaCl$_6$ samples are air-sensitive, they were sealed in a quartz capillary together with helium gas. The measurements were performed at 10 and 300 K.

### E. Physical properties measurements

Magnetization was measured by a commercial SQUID magnetometer (MPMS, Quantum Design). Specific heat measurements are performed by a relaxation method in commercial apparatus (PPMS, Quantum Design).

### F. Nuclear magnetic resonance measurements

Nuclear magnetic resonance (NMR) measurements were carried out with a standard coherent pulsed spectrometer. To determine Knight shift, the gyromagnetic ratio for $^{133}$Cs of 5.5843794 MHz/T was used. Because of the cubic symmetry, the quadrupole splitting of $^{133}$Cs NMR line (nuclear spin $I$ = 7/2) was negligibly small (Fig. 2(c)). Due to the large electric field gradients for the less symmetric Cl site, the $^{35}$Cl NMR ($I$ = 3/2) spectrum is broadened by second-order splitting (Fig. 3(c)). The spin-lattice relaxation rate $1/T_1$ was obtained by fitting the recovery curve of the spin-echo intensity $I(t)$ with the formula for the transition between $I$ = ±1/2 states of $I$ = 3/2 spin: $I(t) = I_{eq} + I_0[0.1\exp(-t/T_1) + 0.9\exp(-6t/T_1)]$ [26].

## III. RESULTS

### A. $J_{eff}$ = 3/2 state and suppression of magnetic dipolar moment in Cs$_2$TaCl$_6$ and Rb$_2$TaCl$_6$

The crystal structures of the polycrystalline samples of Cs$_2$TaCl$_6$, Rb$_2$TaCl$_6$ and K$_2$TaCl$_6$ are confirmed to be cubic at room temperature with a lattice constant $a$ = 10.3285(4) Å, 10.0686(4) Å and

9.9786(5) Å, respectively (see Supplementary Material for details.). The regular shape of the TaCl$_6$ octahedra should give rise to the full degeneracy of the $J_{eff}$ = 3/2 quartet. The small ionic radius of K$^+$, which results in the smallest lattice constant among the three Ta chlorides, leads to a lattice instability and there is a structural phase transition just below room temperature that is discussed below. We therefore focus mainly on Cs$_2$TaCl$_6$ and Rb$_2$TaCl$_6$ in the following.

The realization of the $J_{eff}$ = 3/2 ground state in Rb$_2$TaCl$_6$ is consistent with the local electronic excitation observed in RIXS at the Ta $L_3$-edge. The RIXS spectrum of the powder sample of Rb$_2$TaCl$_6$ (Fig. 1(b)) at 300 K displays a sharp peak at 0.4 eV and a broad one at 3.2 eV, which corresponds to the local $t_{2g}$-$e_g$ excitation. The sharp peak at 0.4 eV can be assigned to an intra-$t_{2g}$ excitation from the degenerate $J_{eff}$ = 3/2 in the cubic symmetry to $J_{eff}$ =1/2 states (inset of Fig. 1(b)). A reasonable magnitude of SOC ($\lambda_{SO}$) of Ta 5$d$ electron ~ 0.27 eV is indeed deduced from the excitation energy from $J_{eff}$ = 3/2 to 1/2: $\Delta E = 3/2 \lambda_{SO}$.

The behavior of magnetic susceptibility, $\chi(T)$, for Cs$_2$TaCl$_6$ and Rb$_2$TaCl$_6$ (Fig. 2(a)) is fully consistent with what is expected for a $J_{eff}$ = 3/2 state, as discussed below. $\chi(T)$ at room temperature is small, of the order of 10$^{-5}$ emu/mol and even diamagnetic for Cs$_2$TaCl$_6$, indicating that 5$d^1$ Ta$^{4+}$ ions in these chlorides are almost non-magnetic. At low temperatures, a small Curie-Weiss contribution is observed. At first glance this looks like the contribution from a small amount of magnetic impurities. However, the good agreement between the temperature dependence of $\chi(T)$ and the Knight shift, $K(T)$, for Cs$_2$TaCl$_6$, obtained from $^{133}$Cs-NMR spectra (Fig. 2(b)) above 10 K, indicates the presence of uniform Curie-Weiss-like contribution to $\chi(T)$. The majority of the Curie-Weiss contribution should come from intrinsic but negligibly small magnetic moments.

The intrinsic Curie-Weiss contribution can be analyzed by fitting the susceptibility with $\chi(T) = \chi_0 + C/(T - \Theta)$, where $C$, $\Theta$ and $\chi_0$ are the Curie-constant, the Curie-Weiss temperature and the temperature independent constant (inset of Fig. 2(a), Table I). Assuming $\chi_0 = \chi_{core} + \chi_{VV}$ and taking the core diamagnetism $\chi_{core}$ from the literature [27], the Van-Vleck susceptibilities $\chi_{VV}$ were estimated for Cs$_2$TaCl$_6$ and Rb$_2$TaCl$_6$ as 1.64 × 10$^{-4}$ and 1.92 × 10$^{-4}$ emu/mol, respectively, and there is a good agreement between the two compounds. This is reasonable as $\chi_{VV}$ originates predominantly from Ta$^{4+}$ ions in almost the same environment. The obtained Curie-constant indicates a $p_{eff}$ of 0.25 $\mu_B$/Ta for Cs$_2$TaCl$_6$ and 0.27 $\mu_B$/Ta for Rb$_2$TaCl$_6$. They are much smaller than those observed in double perovskite oxides and GaTa$_4$Se$_8$. The finite $p_{eff}$ should originate from the uncancelled magnetic dipolar moments and are substantially reduced from the 1.73 $\mu_B$ expected for an $S$ = 1/2 system with $g$ = 2.

The degenerate states can therefore be described by the hidden pseudo-dipolar moments with the multipolar degrees of freedom. The antiferromagnetic Curie-Weiss temperature, $|\Theta|$, is 22 and 30 K for $Cs_2TaCl_6$ and $Rb_2TaCl_6$, respectively, which gives a measure of the energy scale of interaction between the spin-orbital entangled objects. The orderings of multipoles may be expected below 30 K. A pronounced deviation from the Curie-Weiss behavior and anomalies in $\chi(T)$ are indeed observed at low temperatures around 30 K, which reflects the expected multipolar transitions and will be described later.

The substantially reduced $p_{eff}$, as compared with those of $5d^1$ double perovskite oxides and $GaTa_4Se_8$, should reflect the very small hybridization with ligands expected for the chlorides. The effect of hybridization is described by a parameter $\gamma$ defined as $<M> = <2S - \gamma L_{eff}>$, which gives rise to an effective $g$-factor of $2(1 - \gamma)/3$ [27]. The observed effective dipolar moments yield an effective $g_{eff} = 0.13$ (0.14), which corresponds to a $\gamma = 0.81$ (0.79) for $Cs_2TaCl_6$, ($Rb_2TaCl_6$) (Table I), indicating that as much as 80% of the orbital moment is active and compensates the spin moment. We note that, even for $Ba_2NaOsO_6$ with the minimum effective moment among the candidates of $J_{eff} = 3/2$ system known so far, $\gamma$ is as small as $\sim 0.5$ [28].

As seen from Fig. 2(a), a large $p_{eff}$ of 0.96 $\mu_B$/Nb comparable to those of double perovskite oxides, is observed in $Rb_2NbCl_6$ [29], the $4d$ analogue of $Rb_2TaCl_6$. This can be attributed to the increased hybridization and the reduced spin-orbit coupling. Since the Cl-$3p$ level is closer to the Nb-$4d$ level than Ta-$5d$, stronger $d$-$p$ hybridization is expected, which should give rise to an enhanced super-exchange coupling between the $J_{eff} = 3/2$ moments via Cl-$3p$. A large Curie-Weiss temperature $|\Theta| \sim 130$ K (inset of Fig. 2(a)), much larger than those of the $5d$ Ta chlorides $|\Theta| \sim 30$ K, is indeed observed, showing the increased hybridization. The smaller spin-orbit coupling in the $4d$ analogue should reduce the gap between the $J_{eff} = 1/2$ doublet and $J_{eff} = 3/2$ quartet, which enhances the admixture of magnetic $J_{eff} = 1/2$ into $J_{eff} = 3/2$ through the inter-site hopping. The reduced gap between $J_{eff} = 1/2$ and $3/2$ is evidenced by the much larger $\chi_{VV}$ of $4.15 \times 10^{-4}$ emu/mol in $Rb_2NbCl_6$ than in $A_2TaCl_6$ (Table I).

## B. Ordering of multipoles in $Cs_2TaCl_6$ and $Rb_2TaCl_6$

The Curie-Weiss temperature $|\Theta| \sim 20$-$30$ K for the Ta chlorides implies the possible ordering of the pseudo-dipolar moments accompanied with multipolar moments around 30 K. Indeed, we find anomalies in the small Curie-Weiss-like contribution to $\chi(T)$, which reflects the uncanceled magnetic dipolar moments and can be used as a marker for the pseudo-dipolar moments. With decreasing temperature, $\chi(T)$ of Cs$_2$TaCl$_6$ (Fig. 2(a) and the top panel of Fig. 3(a)) and Rb$_2$TaCl$_6$ (Fig. 2(a) and the top panel of Fig. 3(e)) shows a deviation from Curie-Weiss behavior around $T_Q$ = 30 K and 45 K, respectively, followed by a broad peak. Then a clear kink is observed around $T_{PD}$ = 5 K for Cs$_2$TaCl$_6$ and 10 K for Rb$_2$TaCl$_6$, below which a decrease of $\chi(T)$ is observed. The two almost-hidden magnetic anomalies are found to accompany a large total entropy change much larger than the $R$ln2 associated with a doublet and close to $R$ln4, which can be ascribed to lifting of the full degeneracy of the $J_{eff}$ = 3/2 quartet. We therefore argue that these two anomalies represent the ordering of multipolar degrees of freedom attached to the pseudo-dipolar $J_{eff}$ = 3/2 moments.

In the second panel of Fig. 3(a) and (e), we show the specific heat divided by temperature, $C(T)/T$, for Cs$_2$TaCl$_6$ and Rb$_2$TaCl$_6$, where pronounced anomalies are clearly identified at low temperatures. The electronic specific heat divided by temperature, $C_{el}(T)/T$, in the third panel of Fig. 3(a) and (e) is obtained from $C(T)/T$ in the second panel by subtracting the lattice contribution estimated from that of its 5$d^0$ analogue Cs$_2$HfCl$_6$ (for the details of estimation, see the second panels in Fig. 3(a) and (e) and Fig. S2 in Supplementary Material). An appreciable electronic contribution can be seen predominantly below $T = T_Q$ with a small tail extending to above $T_Q$. No clear anomaly in $C_{el}(T)/T$ is identified at $T = T_Q$. At $T \sim T_{PD}$, a clear and sharp peak of $C_{el}(T)/T$ is observed, suggesting that the kink of $\chi(T)$ at $T_{PD}$ corresponds to a well-defined phase transition. The electronic entropy $S_{el}(T)$ was calculated by integrating $C_{el}(T)/T$ (the third panel of Fig. 3(a) and (e)). With increasing temperature, $S_{el}(T)$ increases rapidly up to $T = T_{PD}$ (5 K for Cs$_2$TaCl$_6$ and 10 K for Rb$_2$TaCl$_6$) and then gradually above $T = T_{PD}$. At $T \sim T_Q$ (30 K for Cs$_2$TaCl$_6$ and 45 K for Rb$_2$TaCl$_6$), $S_{el}$ recovers approximately 70% of $R$ln4 and appears to approach gradually to $R$ln4 above $T_Q$. The gradual recovery of $R$ln4 entropy above $T_Q$ can be ascribed to the presence of a short-range ordering (fluctuations) above $T_Q$, which is evident from the lack of a pronounced anomaly at $T_Q$ in $C_{el}(T)$.

We find clear evidences for a structural phase transition from a cubic to a compressed tetragonal at $T = T_Q$, although the thermodynamic signature of transition is very weak, likely due the presence of pronounced fluctuations and short-range ordering. The (004) Bragg peak of x-ray powder diffraction for Cs$_2$TaCl$_6$ in Fig. 3(b) shows a splitting into two peaks at 11 K and 4 K, indicating a phase transition from cubic to tetragonal. The refinement of the powder pattern at 4 K is fully consistent with the space group of $I4/mmm$ and the tetragonal lattice constants $a_T$ = 7.2893(5) Å and $c$ = 10.160(1) Å. Note that $a_T$ corresponds to ~ 1/$\sqrt{2}$ of the cubic lattice constant $a$, which gives a compressive $c/\sqrt{2}a_T$ ratio of 0.985. In the tetragonal phase of Cs$_2$TaCl$_6$, TaCl$_6$ octahedra are uniformly compressed along the $c$-axis, resulting in two distinct Ta-Cl bond lengths of 2.435(15) Å (axial and short) and 2.535(7) Å (equatorial and long) at 4 K (Fig. 3c and Table II). The magnitude of tetragonal distortion $\Delta(c/(\sqrt{2}a_T)$ for $Cs_2TaCl_6$ as a function of temperature (the fourth panel of Fig. 3(a)) shows a pronounced decrease around $T_Q$ = 30 K, evidencing the occurrence of a cubic to tetragonal structural transition at $T = T_Q$. We note that the same compressive-type tetragonal distortion is identified also in the powder XRD measurements of $Rb_2TaCl_6$ at $T$ = 20 K < $T_Q$, where the lattice parameters are $a_T$ = 7.0946(3) Å and $c$ = 9.8072(5) Å with $c/\sqrt{2}a_T$ ratio of 0.977.

The $^{35}$Cl-NMR powder-pattern for $Cs_2TaCl_6$ shown in Fig. 3(d) is fully consistent with the tetragonal distortion below $T_Q$. In the cubic phase, due to a four-fold rotational symmetry at the Cl site, two peaks are expected to appear at the edges of resonance line broadened by the nuclear quadrupolar splitting. This is certainly observed at 40 K in accord with the cubic symmetry. At 10 K, however, each peak at the edge further splits into two peaks (see Fig. 3(d)), consistent with the emergence of two crystallographically different Cl sites at axial and equatorial positions in the tetragonal phase. The phase transition at $T_Q$ = 30 K in $Cs_2TaCl_6$ is identified more clearly as the decrease of the spin-lattice relaxation rate $1/T_1$ from $^{35}$Cl-NMR below $T_Q$, shown in the bottom panel of Fig. 3(a), which is attributed to a change in the low-lying excitations in the ordered phase.

The compressive distortion of $TaCl_6$ octahedra below $T_Q$ should stabilize the quadrupolar $J_{eff}^z = \pm 1/2$ doublet with dominant $xy$-character. As the distortion occurs uniformly, we conclude that the transition at $T_Q$ is a ferro-quadrupolar transition to a $J_{eff}^z = \pm 1/2$ state. The Curie-Weiss temperatures of the uncanceled small magnetic dipolar moments in $\chi(T)$, $|\Theta|$ = 22 K and 30 K for $Cs_2TaCl_6$ and $Rb_2TaCl_6$, respectively, can be viewed as a measure of the energy scale of interaction among the pseudo-dipolar moments and are comparable to and only slightly smaller than $T_Q$ = 30 K and 45 K respectively. This suggests that the primary driving force of the transition at $T_Q$ is the multipolar interactions. It is natural that the coupling with lattice and the Jahn-Teller instability further stabilize the quadrupolar ordering, which may explain the slightly higher $T_Q$ than $|\Theta|$.

The transition at $T_{PD}$ in $Cs_2TaCl_6$ can be ascribed to the antiferro-ordering of $J_{eff}^z = \pm 1/2$ pseudo-dipolar moments with magnetic octupoles. A pronounced broadening of $^{35}$Cl-NMR spectra is observed below $T_{PD}$ as shown in Fig. 3(d), indicative of the appearance of small and inhomogeneous internal magnetic fields, and hence a static ordering of the small and uncanceled magnetic dipolar moments at $T_{PD,}$ to which the octupoles are coupled to. The decrease of $\chi(T)$ with decreasing temperature in Fig. 3(a) without appreciable hysteresis suggests that the ordering of pseudo-dipolar moments is antiferromagnetic. Reflecting the ordering of pseudo-dipoles, the relaxation rate $1/T_1$ shows a sharp decrease by an order of magnitude below $T_{PD}$. The electronic entropy $S_e(T)$ at $T_{PD}$ is ~ 3.5 J K$^{-1}$ mol$^{-1}$, about 60% of $R\ln 2$ both in Cs$_2$TaCl$_6$ and Rb$_2$TaCl$_6$ and reaches to $R\ln 2$ above $T_{PD}$ (~ 10 K and ~ 20 K for Cs and Rb respectively), as shown in Figs. 3(a) and 3(e). This is consistent with the idea that the majority of the $R\ln 2$ entropy of the $J_{eff}^z = \pm 1/2$ doublet is quenched by the ordering of the pseudo-dipoles below $T_{PD}$ and the rest above $T_{PD}$ through the short-range correlations (fluctuations). Apparently, the rest of the $R\ln 4$ entropy of the $J_{eff} = 3/2$ quartet, $R\ln 2$, is quenched by the quadrupolar transition at $T_Q$.

### C. Distinct behavior of K$_2$TaCl$_6$

K$_2$TaCl$_6$, which has a smaller K ion than Cs and Rb ions, has the same cubic structure as Cs$_2$TaCl$_6$ and Rb$_2$TaCl$_6$ at room temperature but exhibits a qualitatively different behavior at low temperatures. The RIXS spectrum at room temperature is consistent with the presence of a degenerate $J_{eff} = 3/2$ ground state (see Supplementary Material).

$\chi(T)$ of K$_2$TaCl$_6$ shown in Fig. 4(a) behaves like those of almost non-magnetic Cs$_2$TaCl$_6$ and Rb$_2$TaCl$_6$ around room temperature. With a $p_{eff}$ of 0.30 $\mu_B$/Ta comparable to those of Cs$_2$TaCl$_6$ and Rb$_2$TaCl$_6$, $\Theta$ of 30 K and $\chi_0$ of -1.46 × 10$^{-5}$ emu/mol are obtained from the Curie-Weiss fitting of $\chi(T)$ in the temperature range between 280 K and 350 K (inset of the top panel of Fig. 4(a)). The $\chi_0$ gives $\chi_{VV}$ of 1.70 × 10$^{-4}$ emu/mol, which is comparable to those of Cs$_2$TaCl$_6$ and Rb$_2$TaCl$_6$. This may justify the Curie-Weiss fitting for K$_2$TaCl$_6$, which was performed in the much narrower temperature range than those for Cs$_2$TaCl$_6$ and Rb$_2$TaCl$_6$.

A well-defined kink in $\chi(T)$ is observed at $T_S = 280$ K (inset of the bottom panel of Fig. 4(a)), which is much higher than those of Cs$_2$TaCl$_6$ and Rb$_2$TaCl$_6$ and, more importantly, than the Curie-Weiss temperature $\Theta$ of 30 K. This means that the transition at $T_S$ likely has a distinct character from those of the $T_Q$ transitions in Cs$_2$TaCl$_6$ and Rb$_2$TaCl$_6$ and that factors other than the interactions between the pseudo-dipolar moments promote the transition at $T_S$. On further cooling, $\chi(T)$ exhibits a pronounced anomaly at $T_M = 15$ K, below which the system is weakly ferromagnetic. In the magnetization curve at 2 K, we observe the typical behavior of a weak ferromagnet and a small saturation moment of ~ 0.02 $\mu_B$/Ta and hysteresis (Fig. 4(b)). This is again in contrast with the antiferro-ordering of uncompensated magnetic moments below $T_{PD}$ in Cs$_2$TaCl$_6$ and Rb$_2$TaCl$_6$.

Powder and single-crystal XRD measurements on K$_2$TaCl$_6$ revealed the presence of a structural phase transition from cubic to tetragonal at $T_S$ = 280 K, as demonstrated by the splitting of the cubic (004) peak in Fig. 4(c). The tetragonal structure below $T_S$ has the space group $P4/mnc$ and lattice constants $a_T$ = 6.8495(3) Å and $c$ = 10.2476(5) Å at 100 K. Below $T_S$, in sharp contrast to the $Cs_2TaCl_6$ and $Rb_2TaCl_6$, the $TaCl_6$ octahedra in the tetragonal phase of $K_2TaCl_6$ are *elongated* along the $c$-axis, with two distinct axial and equatorial Ta-Cl bond lengths of 2.423(4) and 2.381(3) Å at 100 K (Fig. 4(d) and Table II). The magnitude of the tetragonal distortion, $\Delta(c/(\sqrt{2}a_T))$, for $K_2TaCl_6$ as a function of temperature (the second panel of Fig. 4(a)) increases from $T_S$ ~ 280 K to ~ 4% at 4 K (~ 4% elongation). This implies the stabilization of the $J_{eff}^z = \pm 3/2$ states rather than $J_{eff}^z = \pm 1/2$ state below $T_S$ in $K_2TaCl_6$. In addition, the octahedra in the tetragonal phase of $K_2TaCl_6$ are rotated around the $c$-axis by ~10° at 100 K (Fig. 4(d)). The octahedra rotate uniformly in each $ab$-plane, and the direction of rotation alternates along the $c$-axis. It is natural that $J_{eff}^z = \pm 3/2$ states stabilized in the presence of the tensile and rotational tetragonal distortions give rise to a distinct pseudo-dipolar ordering at $T_M$ from those of $Cs_2TaCl_6$ and $Rb_2TaCl_6$.

We argue that the emergence of a distinct tetragonal phase in $K_2TaCl_6$ is the consequence of size mismatch between $K^+$ and $Cl^-$ ions as often observed in isostructural halide compounds [30]. The crystal structure of $A_2TaCl_6$ consists of the close-packed layers made of $A^+$ and $Cl^-$ ionic spheres stacked along the (111) direction, of which octahedral voids are partially filled by $Ta^{4+}$ ions. The ionic radius of $Cl^-$ (1.81 Å) is similar to those of $Rb^+$ (1.72 Å) and $Cs^+$ (1.88 Å) but much larger than that of $K^+$ (1.64 Å) [31]. The large ionic size mismatch in $K_2TaCl_6$ may destabilize the cubic structure and brings about the qualitatively different tetragonal distortion at much higher temperature, independent of pseudo-dipolar interactions.

We note that the $K_2TaCl_6$-type tetragonal distortion at low temperatures, with the alternative rotation of octahedra along the $c$-axis, is common in the family of FCC halides with large ion-size mismatch, and is observed even in a band insulator $Rb_2SnI_6$ without active $d$-electrons [30]. This is consistent with that the $T_S$ transition in $K_2TaCl_6$ being predominantly structural in nature. Interestingly, $Cs_2SnI_6$, a Cs analogue of $Rb_2SnI_6$ with a smaller ion-size mismatch, remains cubic down to the lowest temperature measured, which may be in favor of the $T_Q$ transitions in $Cs_2TaCl_6$ and $K_2TaCl_6$ being driven by electric quadrupole interactions.

## IV. DISCUSSION

It may be informative to compare the results for $Cs_2TaCl_6$ and $Rb_2TaCl_6$ with the theoretically proposed phase diagram for $5d^1$ FCC lattice systems. The theoretical work of Chen et al. [5] determined the ground-state phase diagram as a function of the quadrupolar interaction, the ferromagnetic exchange interaction and the antiferromagnetic exchange interaction, and also the evolution of the ordered phases as a function of temperature. In the presence of a sizable electric quadrupolar interaction, a quadrupolar transition occurs first on lowering the temperature over a wide region of the parameter space. On further reduction of the temperature, the strength of the ferromagnetic interactions determines whether the quadrupolar state has a first-order transition to an antiferro-pseudo-dipolar state with $J_\text{eff}^z = \pm 3/2$, or a second-order transition to a (110) ferro-pseudo-dipolar state consisting of a superposition of $J_\text{eff}^z = \pm 1/2$ and $J_\text{eff}^z = \pm 3/2$ and accompanied by magnetic octupoles.

Our observation of high temperature quadrupolar ordering may indicate the presence of sizable quadrupolar interactions in $Cs_2TaCl_6$ and $Rb_2TaCl_6$. The observed ferro-quadrupolar order appears different from the staggered quadrupolar order discussed in Ref. [5], while the presence of minor staggered components cannot be excluded because the corresponding lattice distortion may be too small to be detected in our XRD measurements. We also observe a pseudo-dipolar ordering at a lower temperature $T_\text{PD}$ as in the theoretical results. The observed behavior of the pseudo-dipolar ordered state is qualitatively different from the two theoretically proposed ground states [5], the antiferro-pseudo-dipolar state and the (110) ferro-pseudo-dipolar state. In the theory, the transition from the quadrupolar state to the antiferro-pseudo-dipolar state with $J_\text{eff}^z = \pm 3/2$ occurs as a first-order transition, accompanied with an elongated octahedra. In the present experiments, the octahedra remains compressed below $T_\text{PD}$ with the difference of axial and equatorial Ta-Cl bond lengths of $0.1 \pm 0.01$ Å (Table II). This excludes the possibility of transition to the antiferro-pseudo-dipolar state with $J_\text{eff}^z = \pm 3/2$ at $T_\text{PD}$ in $Cs_2TaCl_6$ and $Rb_2TaCl_6$. The theoretically predicted transition to the (110) ferro-pseudo-dipolar state is a second order transition, accompanied by increased mixing of a $J_\text{eff}^z = \pm 3/2$ component with lowering temperature. If it were the case, in the experiment below $T_\text{PD}$, a reduction of the compression of octahedra would be observed to stabilize the $J_\text{eff}^z = \pm 3/2$ component, as well as a weak ferromagnetic moment originating from uncanceled dipolar moments. We observe neither the appearance of the ferromagnetic moment nor the suppression of the structural compression below $T_\text{PD}$ ~ 5 K, which excludes also the possibility of (110) ferro-pseudo-dipolar state in $Cs_2TaCl_6$ and $Rb_2TaCl_6$.

We suggest that the coupling of quadrupoles with the lattice plays a crucial role here. The compression of octahedra below $T_Q$ in $Cs_2TaCl_6$ and $Rb_2TaCl_6$ should stabilize the $J_\text{eff}^z = \pm 1/2$ state further energetically. This may suppress the mixing of or the switching to the $J_{\text{eff}}^z = \pm 3/2$ state, stabilizing the ferro-quadrupolar order and constraining the pseudo-dipolar ordering within the $J_{\text{eff}}^z = \pm 1/2$ subspace. Such a situation is treated separately as the easy-plane anisotropy limit in the Ref. [5]. The ground state is predicted to be antiferro-pseudo-dipolar ordering of $J_{\text{eff}}^z = \pm 1/2$ state. The ground state of the $S = 1/2$ FCC lattice model with uniform tetragonal distortion, which may describe the ordering pattern of $J_{\text{eff}}^z = \pm 1/2$ pseudo-dipolar moments in $Cs_2TaCl_6$, is studied in Ref. [32]. Phases consisting of the combination of ferro/antiferromagnetic ordering within the $xy$-plane, with spins pointing within the $xy$-plane or along $z$-axis stacked ferro/antiferromagnetically along the $z$-axis are predicted depending on the sign and magnitude of the relevant interactions. From the broad NMR powder pattern alone, however, we could not determine the ordering pattern of $J_{\text{eff}}^z = \pm 1/2$ pseudo-dipolar moments experimentally. In the case of $K_2TaCl_6$, as discussed above, the structural phase transition forced by the mismatch of ionic size of Cl and K ion stabilizes the $J_{\text{eff}}^z = \pm 3/2$ states. This should favor the theoretically proposed phases dominated by $J_{\text{eff}}^z = \pm 3/2$, and may stabilize the (110) ferromagnetic state. The presence of a rotation of the octahedra may in addition play a role in determining the ordered structure of pseudo-dipolar moments.

## V. CONCLUSION

We discovered that a family of $5d^1$ chlorides $A_2TaCl_6$ ($A$ = Cs, Rb, K) is an ideal arena to explore the exotic state of spin-orbital entangled objects. With the interplay of strong spin-orbit coupling of 0.3 eV, symmetric octahedral environment and strong ionicity, an ideal $J_{\text{eff}} = 3/2$ state of $5d^1$ electrons is realized in these chlorides. We have shown that these chlorides realize a system with significantly suppressed magnetic dipolar moments and that the degeneracy of the $J_{\text{eff}} = 3/2$ quartet is therefore characterized by the pseudo-dipolar moments with the charge quadrupoles and magnetic octupoles. These spin-orbital entangled objects were found to interact with each other weakly on an energy scale of a few tens of kelvin for $Cs_2TaCl_6$ and $Rb_2TaCl_6$. In $Cs_2TaCl_6$ and $Rb_2TaCl_6$, the weak interaction gives rise to a charge quadrupolar transition with pronounced fluctuations at $T_Q \sim 30$ K and $\sim 45$ K and a pseudo-dipolar (magnetic octupolar) transition at $T_{PD} \sim 5$ and 10 K, respectively. The transition at $T_Q$ is accompanied with a structural phase transition to a tetragonal phase, indicative of ferro-ordering of $J_{\text{eff}}^z = \pm 1/2$ quadrupoles. The ordered state is qualitatively different from those theoretically proposed, very likely due to the coupling with the lattice. In $K_2TaCl_6$, a lattice instability lifts the degeneracy of the $J_{\text{eff}} = 3/2$ quartet, which appears to give rise to a distinct pseudo-dipolar ordering from those of $Cs_2TaCl_6$ and $Rb_2TaCl_6$.

$A_2TaCl_6$ is thus far the most ideal $J_{eff} = 3/2$ electron system, with well-defined multipolar orderings. Further studies on the ordered structure and the dynamics of multipoles would significantly advance the understanding of the physics of spin-orbital entangled matters. For example, unveiling elementary excitations from charge quadrupoles and pseudo-dipolar moments (magnetic octupoles) by inelastic neutron scattering and RIXS experiments above and below the ordering is extremely interesting as exotic excitations have been theoretically predicted [33,34]. Our results also demonstrated that controlling the ionicity of the ligand is the key chemistry to explore the novel electronic phases of spin-orbital entangled objects. Material exploration of other halide families, such as more ionic fluorides, would be of interest.

TABLE I. Effective magnetic moment ($p_{eff}$), Curie-Weiss temperature ($\Theta$), effective $g$-value ($g_{eff}$), temperature-independent contribution to magnetic susceptibility $\chi(T)$ ($\chi_0$), Van-Vleck contribution to $\chi(T)$ ($\chi_{VV}$), and cubic lattice constant ($a$) at 300 K are listed for $Cs_2TaCl_6$, $Rb_2TaCl_6$, $K_2TaCl_6$ and $Rb_2NbCl_6$. $g_{eff}$ was obtained from the relation $p_{eff} = g_{eff}\sqrt{J(J+1)}$ and $J = 3/2$. $\chi_0$ was obtained from the Curie-Weiss fitting of magnetic susceptibility $\chi(T)$. $\chi_{VV}$ is obtained from $\chi_0 = \chi_{core} + \chi_{VV}$, where the core magnetic susceptibility $\chi_{core}$ is calculated from those of constituent ions in Ref. [26].

|  | $p_{eff}$ ($\mu_B$/Ta) | $\Theta$ (K) | $g_{eff}$ | $\chi_0$ (emu/mol) | $\chi_{VV}$ (emu/mol) | $a$ (Å) at 300 K |
|---|---|---|---|---|---|---|
| $Cs_2TaCl_6$ | 0.25 | -22 | 0.13 | $-6.07 \times 10^{-5}$ | $1.64 \times 10^{-4}$ | 10.3285(4) |
| $Rb_2TaCl_6$ | 0.27 | -30 | 0.14 | $-7.47 \times 10^{-6}$ | $1.92 \times 10^{-4}$ | 10.0686(4) |
| $K_2TaCl_6$ | 0.30 | -31 | 0.15 | $-1.46 \times 10^{-5}$ | $1.70 \times 10^{-4}$ | 9.9786(5) |
| $Rb_2NbCl_6$ | 0.96 | -131 | 0.50 | $2.21 \times 10^{-4}$ | $4.15 \times 10^{-4}$ | 10.0740(4) |

TABLE II. Tetragonal lattice parameters $a_T$ and $c$, and two Ta-Cl bond lengths, Ta-$Cl_{ax}$ and Ta-$Cl_{eq}$, within $TaCl_6$ octahedron in the low temperature tetragonal phase of $Cs_2TaCl_6$, $Rb_2TaCl_6$ and $K_2TaCl_6$, at a temperature shown in the second column, are listed. $Cl_{ax}$ and $Cl_{eq}$ represent axial and equatorial Cl atoms of the $Cl_6$ octahedron.

|  | $T$ (K) | $a_T$ (Å) | $c$ (Å) | Ta-$Cl_{ax}$ (Å) | Ta-$Cl_{eq}$ (Å) |
|---|---|---|---|---|---|
| $Cs_2TaCl_6$ | 4 | 7.2893(5) | 10.160(1) | 2.435(15) | 2.535(7) |
| $Rb_2TaCl_6$ | 20 | 7.0946(3) | 9.8072(5) | 2.459(4) | 2.543(3) |
| $K_2TaCl_6$ | 100 | 6.8495(3) | 10.2476(5) | 2.423(4) | 2.381(3) |

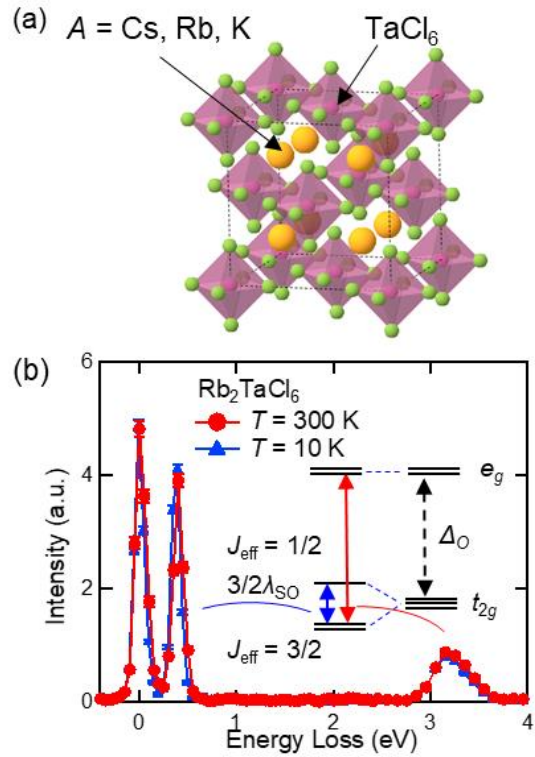

FIG. 1. (a) Crystal structure of $A_2TaCl_6$. The isolated $TaCl_6$ octahedra are colored with purple. $A$ and Cl atoms are shown by yellow and light green spheres. (b) Ta $L_3$-edge RIXS spectrum of the powder sample of $Rb_2TaCl_6$ measured at 300 K (red) and 10 K (blue). The energy diagram of splitting of $d$-orbitals in the cubic octahedral crystal field $\Delta_O$ with and without spin-orbit coupling $\lambda_{SO}$ is shown in the inset.

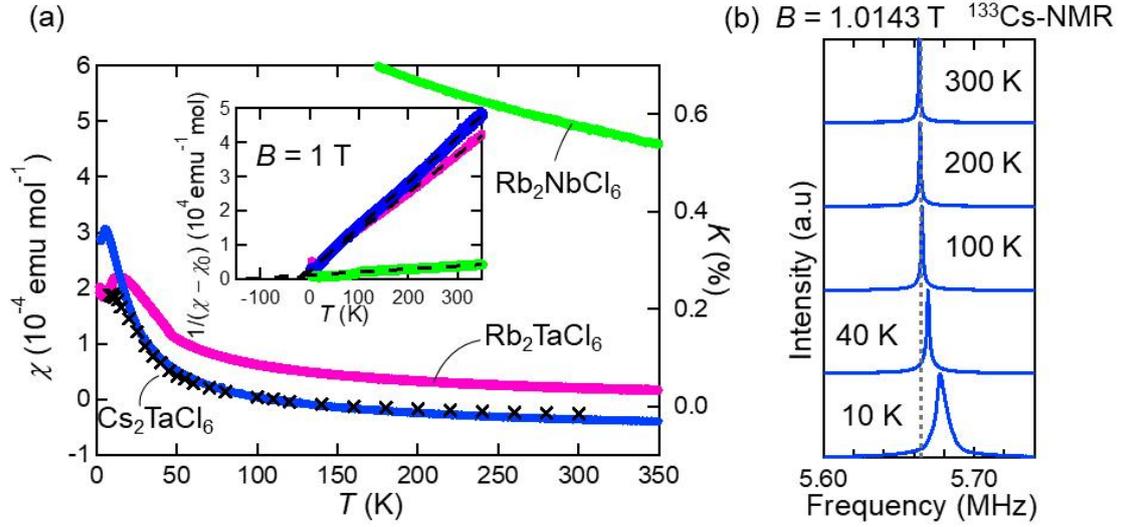

FIG. 2. (a) Magnetic susceptibility $\chi(T)$ of $Cs_2TaCl_6$ (blue), $Rb_2TaCl_6$ (magenta), and $Rb_2NbCl_6$ (green) at $B = 1$ T. The black crosses indicate the Knight shift $K$ of $Cs_2TaCl_6$ obtained from the NMR spectra in (b). The inset shows the plot of the inverse magnetic susceptibility $1/(\chi - \chi_0)$, where $\chi_0$ is a constant obtained from the Curie-Weiss fitting given in Table 1. The intercept of the linear fit (dashed lines) with horizontal axis gives the Curie-Weiss temperatures listed in Table I. (b) $^{133}$Cs-NMR spectra measured at $B = 1.0143$ T by sweeping the frequency at different temperatures. The dashed vertical line indicates zero shift.

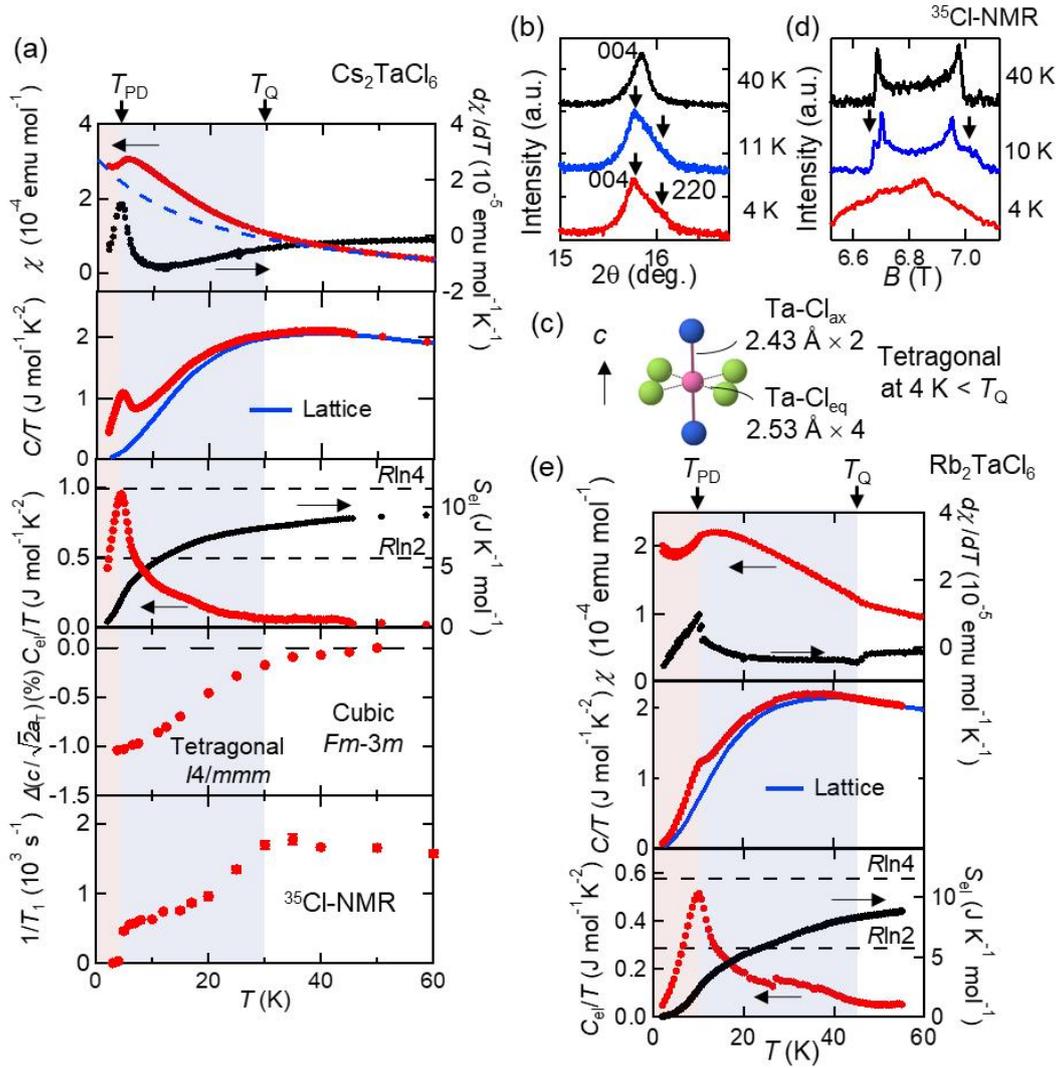

FIG 3. (a) Summary of temperature dependence of various physical parameters characterizing the phase transitions in $Cs_2TaCl_6$. Magnetic susceptibility $\chi(T)$ and its temperature derivative $d\chi/dT$ (the top panel), temperature dependence of specific heat divided by temperature $C/T$ (red) and its lattice contribution (blue) estimated from $C/T$ of $Cs_2HfCl_6$ as discussed in the main text (the second panel), electronic specific heat $C_{el}/T$ (red) and entropy $S_{el}$ (black) (the third panel), tetragonal distortion $\Delta(c/\sqrt{2}a_T)$ (the fourth panel), and spin-lattice relaxation rate $1/T_1$ obtained from the $^{35}Cl$-NMR measurements (the bottom panel). In the third panel, $C_{el}/T$ is obtained by subtracting the lattice estimated contribution from $C/T$ in the second panel. $S_{el}$ is calculated by integrating $C_{el}/T$. $\Delta(c/\sqrt{2}a_T)$ in the fourth panel is the difference between $c/\sqrt{2}a_T$ at each temperature and $c/\sqrt{2}a_T$ at 50 K, where $c$ and $a_T$ are the tetragonal lattice parameters obtained by the refinement of the powder XRD data assuming the tetragonal structure. $T_Q$ and $T_{PD}$ indicate the transition temperature for charge

quadrupolar and pseudo-dipolar orderings, respectively. (b) Selected Bragg reflections in the powder XRD pattern for $Cs_2TaCl_6$ recorded with Mo-$K\alpha_1$ radiation. (c) The local distortion of a $TaCl_6$ octahedron at 4 K in $Cs_2TaCl_6$. $Cl_{ax}$ and $Cl_{eq}$ represent the axial- and equilateral Cl anions of distorted $TaCl_6$ octahedron, respectively. (d) The central transition line of $^{35}Cl$-NMR spectra measured in magnetic field sweeping at NMR frequency of 28.33 MHz (right). (e) Summary of temperature dependence of various physical parameters characterizing the phase transitions in $Rb_2TaCl_6$. $\chi(T)$ and $d\chi/dT$ (the top panel), $C/T$ and its lattice contribution estimated from the specific heat of $Cs_2HfCl_6$ (the middle panel), and $C_{el}/T$ and entropy $S_{el}$ (the bottom panel).

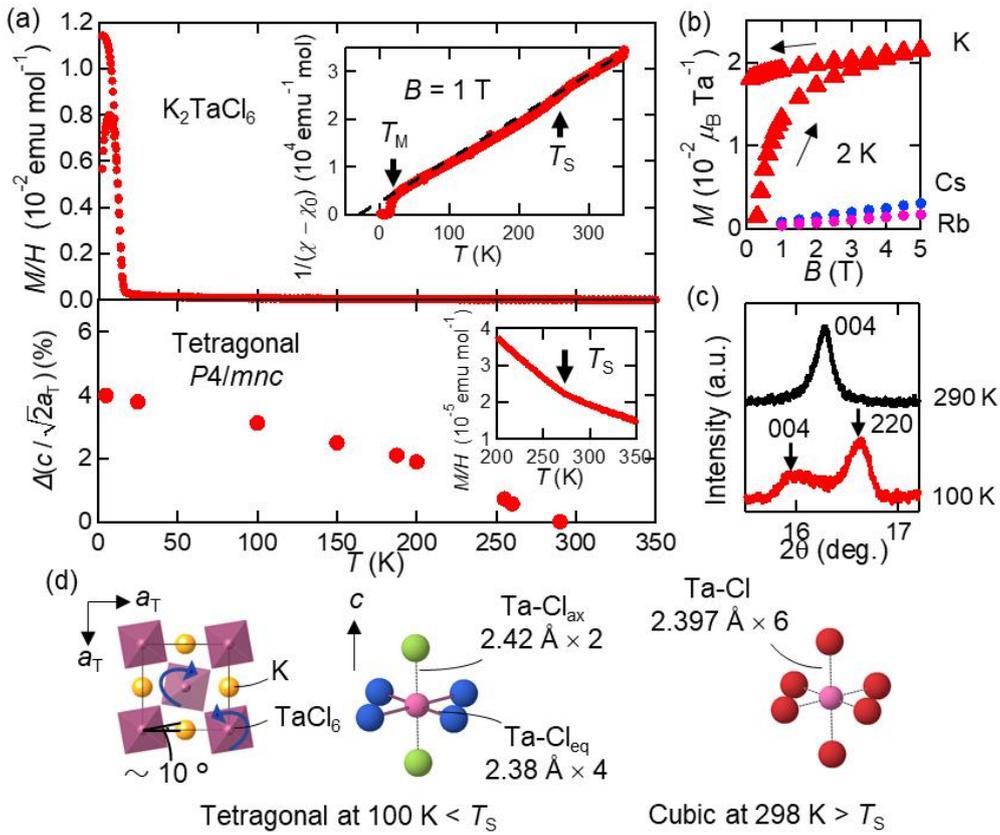

FIG 4. (a) Temperature dependence of magnetization divided by magnetic field, $M/H$, (the top panel) and tetragonal distortion $\Delta(c/\sqrt{2}a_T)$ (the bottom panel) for $K_2TaCl_6$. $\Delta(c/\sqrt{2}a_T)$ is the difference between $c/\sqrt{2}a_T$ at each temperature and $c/\sqrt{2}a_T$ at 290 K, where $c$ and $a_T$ are the tetragonal lattice parameters obtained by the refinement of the powder XRD data assuming the tetragonal structure. The inset to the top panel shows the plot of the inverse magnetic susceptibility $1/(\chi - \chi_0)$, where $\chi_0$ is a constant obtained from the Curie-Weiss fitting given in Table I. The inset to the bottom panel shows

the magnetic susceptibility data enlarged around 280 K. $T_S$ and $T_M$ indicate the temperatures for the structural and the magnetic phase transitions, respectively. (b) Magnetization curves of $K_2TaCl_6$ (red), $Cs_2TaCl_6$ (blue) and $Rb_2TaCl_6$ (magenta) measured at 2 K. (c) Selected Bragg reflections in the powder XRD pattern of $K_2TaCl_6$ at 296 and 100 K recorded with Mo-$K\alpha_1$ radiation. (d) Crystal structure of $K_2TaCl_6$ at 100 K viewed from the $c$-axis determined by the single crystal x-ray diffraction, emphasizing the rotation of $TaCl_6$ octahedra (left). The local distortion of $TaCl_6$ octahedron at 100 K (middle) and the regular $TaCl_6$ octahedron at 298 K (right).

## ACKNOWLEDGEMENTS

We thank George Jackeli, Andrew Smerald, and Leon Balents for stimulating discussions. The synchrotron radiation experiments were performed at the BL11XU of SPring-8 with the approval of the Japan Synchrotron Radiation Research Institute (JASRI) (Proposal No. 2017A3552). This work was partly supported by Japan Society for the Promotion of Science (JSPS) KAKENHI (No. JP15H05852, JP15K21717, 17H01140), JSPS Core-to-core program "Solid-state chemistry for transition-metal oxides", and Alexander von Humboldt foundation.

# Ordering of hidden multipoles in spin-orbital entangled $5d^1$ Ta chlorides

H. Ishikawa,[1,2] T. Takayama,[1,2] R. K. Kremer,[2] J. Nuss,[2] R. Dinnebier,[2] K. Kitagawa,[3] K. Ishii,[4] and H. Takagi[1,2,3]

[1] Institute for Functional Matter and Quantum Technologies, University of Stuttgart, Stuttgart 70569, Germany

[2] Max-Planck-Institute for Solid State Research, Heisenbergstraße 1, Stuttgart 70569, Germany

[3] Department of Physics, The University of Tokyo, 7-3-1 Hongo, Bunkyo-ku, Tokyo 113-0033, Japan

[4] Synchrotron Radiation Research Center, National Institutes for Quantum and Radiological Science and Technology, Sayo, Hyogo 679-5148, Japan

## A. X-ray absorption and RIXS spectra at Ta $L_3$-edge for $K_2TaCl_6$ and $Rb_2TaCl_6$

Figure S1(a) shows the x-ray absorption spectra (XAS) of $Rb_2TaCl_6$ and $K_2TaCl_6$ at Ta $L_3$-edge collected by fluorescence mode at room temperature. The peak of white line appears around 9.879 keV in both samples. Figure S1(b) presents the incident x-ray energy dependence of RIXS spectra for $K_2TaCl_6$ at room temperature. The measurements were performed with a low-energy resolution setup using Si(444) channel-cut monochromator as a secondary monochromator (the total energy resolution was ~ 150 meV). In addition to the elastic peak centered at 0 eV, there are two prominent features at 0.4 eV and 3.2 eV. The latter peak becomes stronger when the incident energy is tuned to the peak of XAS, indicating that it corresponds to the local $t_{2g}$ to $e_g$ excitation of Ta $5d$ electron (10Dq). On the other hand, by reducing the incident energy about 3-4 eV from the peak energy ~ 9.880 keV, the low-energy feature around 0.4 eV is enhanced. This suggests that the 0.4 eV peak represents an excitation within the $t_{2g}$ manifold, most likely originating from the $J_{eff} = 3/2$ to $J_{eff} = 1/2$ excitations.

The high-resolution RIXS measurements on $Rb_2TaCl_6$ and $K_2TaCl_6$ were performed with the incident x-ray energy of 9.877 keV where the excitation peak at 0.4 eV becomes most pronounced as discussed above. The results at 10 K and 300 K are displayed in Fig. 1(b) ($Rb_2TaCl_6$) and Fig. S1(c) ($K_2TaCl_6$). Any appreciable increase of energy of 0.4 eV peak and the emergence of a new low energy peak, expected for the splitting of $J_{eff} = 3/2$ manifolds, are not observed in the spectrum at 10 K (< $T_Q$) as compared with that at 300 K (> $T_Q$), for both Rb and K systems. This means that the splitting of $J_{eff} = 3/2$ manifolds at 10K (< $T_Q$) should be below our energy resolution of ~ 90 meV. The small splitting of $J_{eff} = 3/2$ manifolds is consistent with the small energy scale of interactions among the multipoles,

20-30 K (a few meV), which is estimated from the Curie-Weiss temperature in Fig. 3(a).

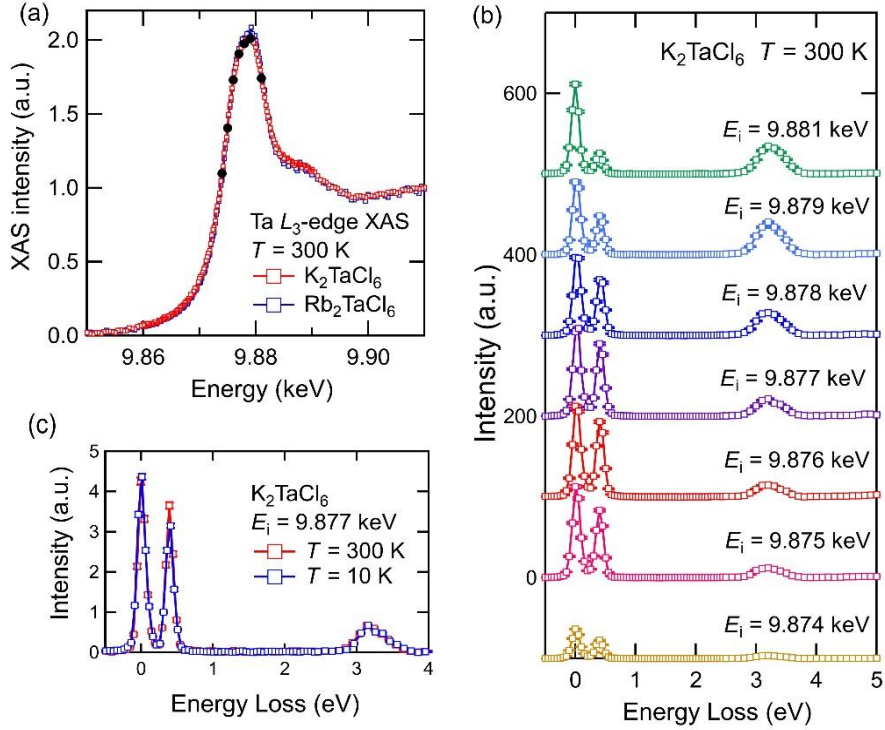

FIG. S1. (a) Ta $L_3$-edge x-ray absorption spectra (XAS) for $K_2TaCl_6$ and $Rb_2TaCl_6$ measured at room temperature. The spectra are normalized with high-energy intensity above 9.90 keV. The black dots on the data of $K_2TaCl_6$ indicate the energy points where the RIXS spectra in b were collected. (b) Ta $L_3$-edge RIXS spectra of $K_2TaCl_6$ measured with different incident energy $E_i$. (c) Ta $L_3$-edge RIXS spectra of $K_2TaCl_6$ with incident energy $E_i$ = 9.877 eV measured at 10 K and 300 K.

### B. Details of estimation of electronic specific heat

Figure S2(a) shows the temperature dependence of specific heat divided by temperature, $C/T$, for $Cs_2TaCl_6$. In order to estimate the lattice contribution to specific heat, $C/T$ of $5d^0$ analogue $Cs_2HfCl_6$ was also measured. $C/T$ of $Cs_2HfCl_6$ is slightly larger than that of $Cs_2TaCl_6$ above 20 K, likely reflecting the subtle difference of the phonon spectra between the two compounds. To compensate the difference, we multiplied a calibration factor to $C/T$ of $Cs_2HfCl_6$ so that it agrees well with that of $Cs_2TaCl_6$ between 50 and 100 K. In this case, a factor of 0.96 appears to work well. The electronic specific heat of $Cs_2TaCl_6$ in Fig. 3(a) of the main text was estimated by subtracting thus obtained lattice contributions from the total $C/T$.

The electronic entropy $S_{el}$, calculated by integrating the electronic specific heat, is shown in Fig. S2(b). We show three different curves for $S_{el}$, obtained by using the different calibration factors for the estimate of lattice contribution. When the factor is larger than 0.96, e.g. 0.98 (black), $S_{el}$ starts decreasing already above 40 K, which is unphysical. The choice of a calibration factor of 0.96 (red) or slightly smaller value, for example, 0.94 (blue) seems to be reasonable. We do not have enough resolution to distinguish 0.94 and 0.96 considering the given uncertainty in the process of estimating the lattice contribution from $Cs_2HfCl_6$. However, we note that the estimate of electronic entropy below 20 K does not depend appreciably on the choice of calibration factor. Already at 20 K, $S_{el}$ exceeds $R\ln2$ and is gradually approaching to $R\ln4$ with increasing temperature. The estimate of electronic entropy of $R\ln2$ right above $T_{PD}$ and close to $R\ln4$ above $T_Q$ is robust. For the estimate of the electronic specific heat in $Rb_2TaCl_6$ shown in Fig. 3(e), the calibration factor was set to 1 (no adjustment) as $C/T$ of $Rb_2TaCl_6$ and $Cs_2HfCl_6$ almost coincide at around 50 K.

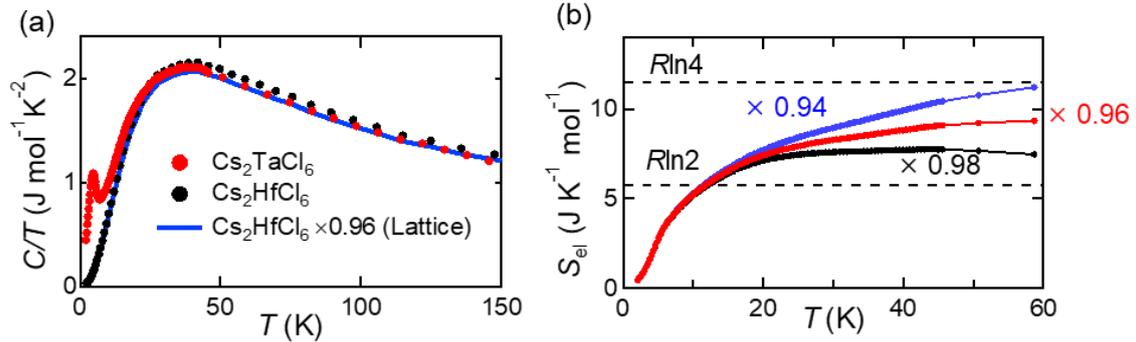

FIG. S2. (a) Temperature dependence of specific heat divided by temperature $C/T$ for $5d^1$ $Cs_2TaCl_6$ (red) and its $5d^0$ analogue $Cs_2HfCl_6$ (black), respectively. The blue curve depicts the estimated lattice contribution obtained by multiplying a factor of 0.96 to $C/T$ of $Cs_2HfCl_6$. (b) The calculated electronic entropy $S_{el}$ with different estimates of lattice contributions. The calibration factor to the specific heat of $Cs_2HfCl_6$ were changed from 0.94 to 0.98 (0.96 is used in Fig. 3(a) and Fig. S2(a)).

$C/T$ for $K_2TaCl_6$ is displayed in Fig. S3(a). The lattice contribution to specific heat was evaluated from $C/T$ of $Cs_2HfCl_6$ as well. To estimate the electronic entropy of $K_2TaCl_6$ below 20 K, a small calibration factor of 0.84 is multiplied to $C/T$ of $Cs_2HfCl_6$ (blue curve). The electronic specific heat divided by temperature $C_{el}/T$, obtained by subtracting the lattice part, is shown in the bottom panel of Fig. S3(b), together with the temperature dependence of magnetization divided by magnetic field $M/H$. The electronic entropy $S_{el}$, calculated by integrating $C_{el}/T$, is close to $R\ln2$ at 20 K, as expected for the pseudo-dipolar ordering of $J_{\text{eff}}^z = \pm 3/2$ state, selected from the $J_{\text{eff}} = 3/2$ quartet below the cubic-tetragonal structural phase transition at $T_S = 280$ K. We note that $T_S = 280$ K is much higher than the magnetic ordering temperature $T_M = 15$ K, in contrast to those of $Cs_2TaCl_6$ and $Rb_2TaCl_6$. The estimate of electronic entropy up to $T_S = 280$ K could not be performed. As the lattice contribution to specific heat becomes dominant at high temperatures above 30 K and it is technically impossible to separate the small electronic contribution.

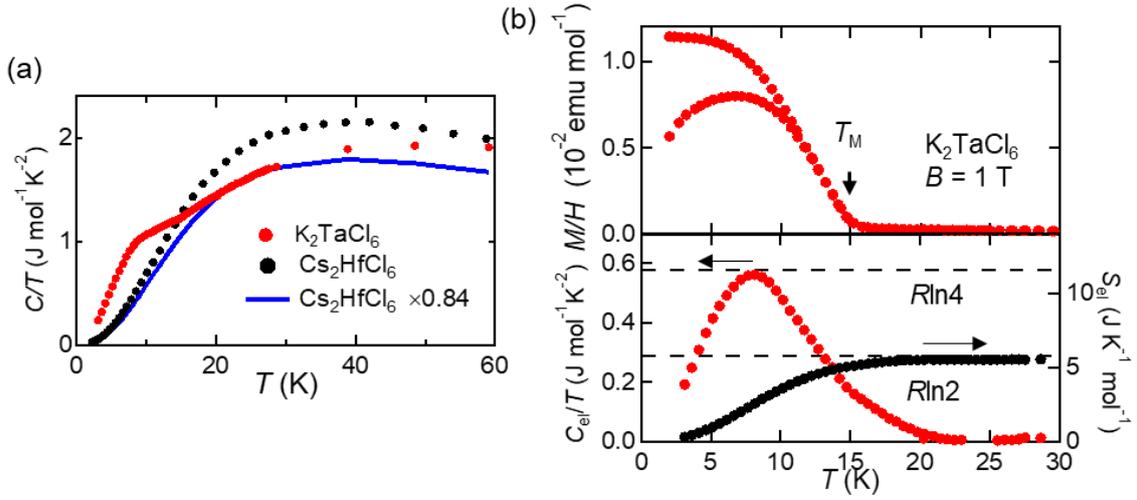

FIG. S3. (a) Specific heat divided by temperature $C/T$ for $K_2TaCl_6$ (red) and $Cs_2HfCl_6$ (black) as a reference for the lattice contribution. The blue curve shows the estimated lattice contribution obtained by multiplying a factor of 0.84 to $C/T$ of $Cs_2HfCl_6$. (b) Magnetization divided by magnetic field $M/H$ for $K_2TaCl_6$ (the top panel). The bottom panel shows the electronic specific heat divided by temperature $C_{\text{el}}/T$ and electronic entropy $S_{\text{el}}$ for $K_2TaCl_6$.

### C. Structural information for $A_2TaCl_6$ ($A$ = Cs, Rb, K).

The lattice constants and atomic coordinates for $A_2TaCl_6$ ($A$ = Cs, Rb, K), refined by the x-ray diffraction measurements, are listed below. The structural parameters were obtained from powder x-ray diffraction for $A$ = Cs and Rb, and from single crystal x-ray diffraction for $A$ = K, respectively. In the tables, the occupancy is unity for all sites. Experimental details of data collection for $A$ = K are also given.

TABLE I. The cubic structure (space group $Fm\text{-}3m$, No. 225) at room temperature for $A_2\text{TaCl}_6$ ($A$ = Cs, Rb, K).

$\text{Cs}_2\text{TaCl}_6$

$a = 10.3285(4)$ Å, $Z = 4$, $T = 298$ K

| Atom | site | $x$ | $y$ | $z$ |
| --- | --- | --- | --- | --- |
| Ta1 | 4$a$ | 0 | 0 | 0 |
| Cs1 | 8$c$ | 0.25 | 0.25 | 0.25 |
| Cl1 | 24$e$ | 0.2319(6) | 0 | 0 |

$\text{Rb}_2\text{TaCl}_6$

$a = 10.0686(4)$ Å, $Z = 4$, $T = 298$ K

| Atom | site | $x$ | $y$ | $z$ |
| --- | --- | --- | --- | --- |
| Ta1 | 4$a$ | 0 | 0 | 0 |
| Rb1 | 8$c$ | 0.25 | 0.25 | 0.25 |
| Cl1 | 24$e$ | 0.2392(5) | 0 | 0 |

$\text{K}_2\text{TaCl}_6$

$a = 9.9778(12)$ Å, $Z = 4$, $T = 298$ K

| Atom | site | $x$ | $y$ | $z$ |
| --- | --- | --- | --- | --- |
| Ta1 | 4$a$ | 0 | 0 | 0 |
| K1 | 8$c$ | 0.25 | 0.25 | 0.25 |
| Cl1 | 24$e$ | 0.24025(11) | 0 | 0 |

TABLE II. The tetragonal structure (space group $I4/mmm$, No. 139) for $A_2\text{TaCl}_6$ ($A$ = Cs and Rb).

$\text{Cs}_2\text{TaCl}_6$

$a = 7.2893(5)$ Å, $c = 10.160(1)$ Å, $Z = 2$, $T = 4$ K

| Atom | site | $x$ | $y$ | $z$ |
| --- | --- | --- | --- | --- |
| Ta1 | 2$a$ | 0 | 0 | 0 |
| Cs1 | 4$d$ | 0 | 0.5 | 0.25 |
| Cl1 | 4$e$ | 0 | 0 | 0.2397(14) |
| Cl2 | 8$h$ | 0.2459(9) | 0.2459(9) | 0 |

Rb$_2$TaCl$_6$

$a$ = 7.0946(3) Å, $c$ = 9.8072(5) Å, $Z$ = 2, $T$ = 20 K

| Atom | site | $x$ | $y$ | $z$ |
|---|---|---|---|---|
| Ta1 | 2$a$ | 0 | 0 | 0 |
| Rb1 | 4$d$ | 0 | 0.5 | 0.25 |
| Cl1 | 4$e$ | 0 | 0 | 0.2507(4) |
| Cl2 | 8$h$ | 0.2535(3) | 0.2535(3) | 0 |

TABLE III. The tetragonal structure (space group $P4/mnc$, No. 128) for $A_2$TaCl$_6$ ($A$ = K)

K$_2$TaCl$_6$

$a$ = 6.8495(3) Å, $c$ = 10.2476(5) Å, $Z$ = 2, $T$ = 100 K

| Atom | site | $x$ | $y$ | $z$ |
|---|---|---|---|---|
| Ta1 | 2$a$ | 0 | 0 | 0 |
| K1 | 4$d$ | 0.5 | 0 | 0.25 |
| Cl1 | 4$e$ | 0 | 0 | 0.2364(3) |
| Cl2 | 8$h$ | 0.1994(3) | -0.2848(3) | 0 |

TABLE IV. Crystal data and structure refinement data of $K_2TaCl_6$ at 298 and 100 K

| Temperature /K | 298 | 100 |
|---|---|---|
| Formula weight | 471.85 | |
| Space group (no.), Z | $Fm\bar{3}m$ (225), 4 | $P4/mnc$ (128), 2 |
| Lattice constants /Å | a = 9.9778(12) | a = 6.8495(3) |
| | | c = 10.2476(5) |
| $V$ /Å$^3$, $\rho_{xray}$ /g cm$^{-3}$ | 993.4(4), 3.155 | 480.77(5), 3.259 |
| Crystal size /mm$^{-3}$ | 0.40×0.30×0.20 | |
| Diffractometer | SMART APEX I, Bruker AXS | |
| X-ray radiation, $\lambda$/Å | 0.71073 | |
| Absorption correction | Multi-scan, SADABS | Multi-scan, TWINABS |
| $2\theta$ range /° | 7.07-70.72 | 7.15-84.57 |
| Index range | $-15 \leq h \leq 15$ | $0 \leq h \leq 9$ |
| | $-15 \leq k \leq 16$ | $0 \leq k \leq 12$ |
| | $-16 \leq l \leq 16$ | $0 \leq l \leq 19$ |
| Reflection collected | 3962 | 22344 |
| Data, $R_{int}$ | 151, 0.035 | 894, 0.080 |
| No. of parameters | 7 | 15 |
| Transmission: $t_{max}, t_{min}$ | 0.108, 0.038 | 0.177, 0.0666 |
| $R_1[F^2 > 2\sigma(F^2)]$ | 0.014 | 0.054 |
| $wR(F^2)$ | 0.032 | 0.140 |
| $\Delta\rho_{max}, \Delta\rho_{min}$ /e Å$^{-3}$ | 0.47, −0.53 | 5.81, −2.92 |
| Deposition no. | CSD-434040 | CSD-434041 |